\documentclass[a4paper,amsmath,amssymb,10pt,aps,pre,twocolumn,floats,floatfix,superscriptaddress,showpacs]{revtex4}
\usepackage{amssymb,epsf}
\usepackage{graphicx}
\usepackage{dcolumn}
\usepackage{bm}
\usepackage{epsfig}
\usepackage{color}

\begin{document}

\title{Scaling of critical connectivity of mobile ad hoc communication networks}
\author{Li Wang}
\affiliation{College of Science, Nanjing University of Aeronautics
and Astronautics, Nanjing, 210016, P. R. China}
\author{Chen-Ping Zhu}\email{chenpingzhu@yahoo.com.cn}
\affiliation{College of Science, Nanjing University of Aeronautics
and Astronautics, Nanjing, 210016, P. R. China}
\affiliation{Department of Physics and Center for Computational
Science and Engineering, National University of Singapore,
Singapore, 117542, Singapore}
\author{Zhi-Ming Gu}
\affiliation{College of Science, Nanjing University of Aeronautics
and Astronautics, Nanjing, 210016, P. R. China}
\author{Shi-Jie Xiong}
\affiliation{Department of Physics and National Laboratory of Solid
State Microstructures, Nanjing University,  Nanjing,  210093,  China
}
\author{ Da-Ren He}
\affiliation{College of Physical Science and Technology, Yangzhou
University, Yangzhou, 225001, P. R. China}
\author{ Bing-Hong Wang}
\affiliation{Department of Modern Physics and Institute of
Theoretical Physics, University of Science and Technology of China,
Hefei, 230026, P. R. China}
\date{\today}

\begin{abstract}
In this paper, critical global connectivity of mobile ad hoc
communication networks (MAHCN) is investigated. We model the
two-dimensional plane on which nodes move randomly with a triangular
lattice. Demanding the best communication of the network, we account
the global connectivity $\eta$ as a function of occupancy $\sigma$
of sites in the lattice by mobile nodes. Critical phenomena of the
connectivity for different transmission ranges $r$ are revealed by
numerical simulations, and these results fit well to the analysis
based on the assumption of homogeneous mixing . Scaling behavior of
the connectivity is found as $\eta \sim f(R^{\beta}\sigma)$, where
$R=(r-r_{0})/r_{0}$, $r_{0}$ is the length unit of the triangular
lattice and $\beta$ is the scaling index in the universal function
$f(x)$. The model serves as a sort of site percolation on dynamic
complex networks relative to geometric distance. Moreover, near each
critical $\sigma_c(r)$ corresponding to certain transmission range
$r$, there exists a cut-off degree $k_c$ below which the clustering
coefficient of such self-organized networks keeps a constant while
the averaged nearest neighbor degree exhibits a unique linear
variation with the degree k, which may be useful to the designation
of real MAHCN.
\end{abstract}
\pacs{89.75.Hc, 89.20.Hh}
\maketitle

Mobile ad hoc communication network (MAHCN)\cite{IETF1,Wireless2} is
a new sort of communication circumstance. It consists of many mobile
nodes carrying out collective duty while nodes communicate with each
other via wireless links. Neither central control authority nor
intermediate services such as base stations for cellular
mobile-phones exist in the network. Each of its nodes needs to relay
packets within its limited transmission range for other participants
in multi-hop edges. MAHCN changes its topology with time without
prior notice since its nodes are free to move
randomly\cite{Perkins3}. Therefore, to realize effective
communication and to do moving jobs, it should self-organize into a
dynamically stationary network by certain local protocols. The study
of mobile ad hoc networks has attracted much attention recently due
to their potential application in battlefield, disaster relief
providing, outdoor assemblies and other settings with temporal,
inexpensive usage. Tens of protocols have been proposed by
designers. However, investigation on property of the connectivity as
viewed from statistical physics is still inadequate, which motivates
the work in the present paper.

 The theory of complex networks\cite{Newman4,Albert5} can provide powerful tools to
investigate the mobile ad hoc networks. Xie \emph{et al.}\cite{Xie6}
analyzed the formation of complex networks involving both geometric
distance and topological degree of nodes. Sarshar \emph{et
al.}\cite{Sarshar7,Sarshar8} noticed that while new nodes are added
to the existing network, other nodes might leave the network rapidly
and randomly. They presented results about the possible emergence of
scale-free structure in ad hoc networks. However, they considered
only static cases instead of analyzing the influence from the motion
of nodes. N\'emeth and Vattay \cite{Nemeth9} studied the giant
cluster of such networks, and pointed out that the giant component
size in the percolation could be described by a single
parameter---the average number of neighbors of nodes. On the other
hand, models of percolation on networks \cite{Newman,Moore} were
often employed to analyze spreading processes, especially epidemics
with occupancy threshold $p_c$ showing drastic transitions.

 A communication network may deliver meaningful services only if the
network is well connected, or at least has a vast subset that is
connected. Therefore, one goal of studying the ad hoc network is to
find out how the network can maintain its connectivity. Different
from previous study\cite{Nemeth9}, we demand global connection of
all nodes in the MAHCN for the best communication, which means a
stricter case than site
percolation\cite{Nemeth9,Newman,Moore,Hu10,Glauche11,Angeles12}. For
simplicity, we model the two-dimensional plane on which nodes move
randomly as a triangular lattice with N vertices, so that we mimic
round transmission ranges with discrete hexagons. We define the
probability of global connection $\eta$ as the ensemble average of
$n/n_0$, where $n$ is the number of moving nodes globally connected
to the integrated network, and $n_0$ is the total number of them.
Critical behavior of order parameter $\eta$ is found to rely on both
transmission range $r$ and the occupancy $\sigma$(defined as
$n_0/N$) of sites (i.e. vertices in the triangular lattice). Scaling
behavior of the order parameter is verified in the form of $\eta
\sim f(R^{\beta}\sigma)$, where $R$ is the reduced transmission
range, and $\beta$ is the scaling index in the universal function
$f(x)$. Moreover, at critical thresholds of occupancy, individual
nodes self-organize into complex networks which display particular
degree distribution\cite{Barabasi}, clustering
coefficient\cite{Watts} and averaged nearest neighbor
degrees\cite{Satorras} of such dynamic communication networks.

  To describe the case of self-organized communication of individual
nodes, it is assumed that they move on a two-dimensional plane in a
discrete way, and a simple dynamic ad hoc network model is proposed
as follows. On the two dimensional triangular lattice(see Fig.1) of
size $L$, individual nodes are assumed to distribute randomly on the
sites of it. In our work, the total number of the sites are chosen
as $N=L^{2}$ with $L=200$. And, under periodic boundary condition,
we restrain the motion of nodes along edges between sites. At the
initial time step, assign $n_0$ nodes to the sites in the triangular
lattice randomly. Every site of such a lattice can be occupied by
only one node or nothing. Considering the dynamic topology of the ad
hoc network, we suppose that, at every time step, each node can move
randomly in one of the six directions to its neighbor site if it is
not occupied. Two nodes in the ad hoc network can communicate with
each other if the distance between them is less than the minimum of
their two transmission ranges\cite{Sanchez17,Santi18}. To simplify
our study, we assume that all the mobile nodes have the same fixed
transmission power\cite{Ramanathan19,Rodoplu20}. Therefore they all
have equal transmission range $r$ for valid communication of the
whole network. The transmission range $r$ is an important parameter
in the designation of ad hoc networks, since it is vital to keep the
network globally connected. Proper adoption of range $r$ could
minimize energy consumption since transmission power is proportional
to the square of it. Note that neither self-loop nor multiple edge
is allowed in the network: (1) A node should not communicate with
itself; and (2) technically there is no sense to open another
communication channel between any two nodes if they are already
neighbors. The motion of nodes at a certain value of occupancy forms
different configurations of site-occupation on the lattice, which
serves as the ensemble for the calculation on the global
connectivity(i.e. the probability of the global connection) and
other averaged quantities.

\begin{center}
\begin{figure}
\scalebox{1.1}[1.1]{\includegraphics{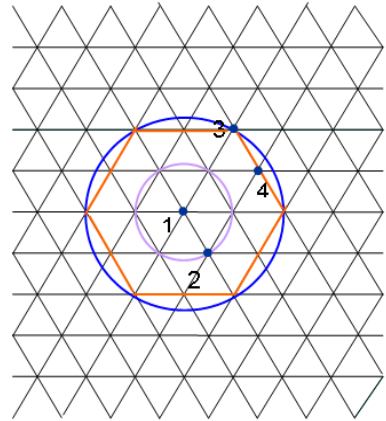}} \caption{(Colour
online)Triangular lattice with transmission range $r=r_0$ and
$r=2r_0$, respectively, where $r_0$ is the length of every edge.}
\end{figure}
\end{center}

  An intuitive way to ensure the best communication by multi-hop
linking is to increase the occupancy of sites on the lattice by
mobile nodes, while this changes the energy consumption of whole the
network. In the framework of traditional percolation problems,
continuously increasing site occupancy will pass threshold of site
percolation. To our knowledge, the term "site"
 can be used dually: one
means a block\cite{Christensen}; the other means a
vertex\cite{Yuge}. Obviously our usage belongs to the later one. Two
sites are neighbors when two occupied vertices are directly linked
by a common bond (edge) or two occupied blocks have a common border.
Therefore, an indirect connection between any two sites means that a
path exists from one to the other neighbor by neighbor. In the
present model, however, every node is located at the center of its
own transmission range $r=zr_0$ (z is a positive integer, and $r_0$
is the length of an edge of any minimal triangle). All the other
nodes at sites inside this circle connect with it directly. Taking
the circle $r=2r_0$ (blue in Fig.1) as an example, nodes 2, 3 and 4
inside the inscribed hexagon (orange in Fig.1) of it are direct
neighbors of node 1. Therefore, neighbors here are determined by the
transmission range and in the sense of topological connection, just
as what occurred in complex network models\cite{Newman,Moore}, which
distinguishes them from those in traditional two-dimensional site
percolation. It is a special case of topological correlation valid
within certain geographic distance\cite{Kim, Holme, Morilta,
Hayashi, Fontoura}. The triangular lattice in our model is a
beneficial setting to describe moving nodes and for discrete
calculation of network parameters. Moreover, in percolation problem
one always focuses on the probability for a site to be included in
the giant component which just extends from a border to its opposite
one in a lattice with finite size. However, for an ad hoc network
bearing search, rescue, tracing or precise attack, the task may
concentrate on a few, even a single moving target. It may demand
global connection of all nodes, which is quite different from the
percolation problem which leaves many nodes scattering outside the
giant component.

  The order parameter $\eta$, i.e. the global
connectivity of the MAHCN, is calculated with burning
algorithm\cite{Herrmann21}. The evolution of it should rely on the
occupancy $\sigma$ of the sites, and $\eta$ enhances when $\sigma$
increases. The ratio of the enhancement depends on the number $n$ of
the nodes which have been connected into the largest dynamic network
at that time step, and it also depends on the number of the
disconnected nodes, based on the assumption of homogeneous
mixing\cite{Anderson} of randomly moving nodes. Therefore, we have

\begin{equation}
\frac{d\eta}{d\sigma} \propto n(n_0-n)
\end{equation}
where $n_0$ is the total number of mobile nodes. Using the
definition of $\eta$ and getting the effect of the transmission
range included, we arrive at

\begin{equation}
\frac{d\eta}{d\sigma}=g(r)\eta(1-\eta)
\end{equation}
where $g(r)$ is the function of transmission range $r$. This
equation can be solved with the uniform initial condition
$\eta(\sigma\to0)=\eta_{0}$:

\begin{equation}
\eta(\sigma)=\frac{\eta_{0}}{\eta_{0}+(1-\eta_{0})e^{-g(r)\sigma}}
\end{equation}

   In fig. 2, simulation results for the
global connectivity as a function of the occupancy of the sites on
the triangular lattice are illustrated. They are in good agreement
with the analytical result of eq.(3) under the condition of
dimensionless function $g(r)\sim{r/{r_0}}$. Actually, it is
naturally expected by dimension analysis on the exponent in the
denominator: $g(r)$  should have no dimension since occupancy
$\sigma$ is dimensionless. The difference between simulation and
analytical results at bottom parts can be attributed to the
deviation from homogenous assumption by the distribution of nodes at
discrete sites on the triangular lattice, and size effect.

\begin{center}
\begin{figure}
\scalebox{0.8}[0.7]{\includegraphics{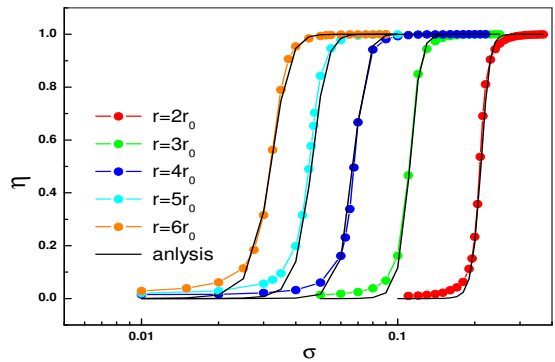}} \caption{(Colour
online)Comparison between theoretical analysis and numerical
simulation on the probability of global connection as a function of
$\sigma$ for various transmission ranges $r$. Hereafter all
simulation results are averaged over 300 realizations of
configurations formed by moving nodes on the triangular lattice.}
\end{figure}
\end{center}

  Numerical results display the critical behavior of
global connectivity, i.e. they show its drastic transitions
occurring at critical values $\sigma_c$ for different transmission
ranges. When occupancy $\sigma$ passes $\sigma_c$, the dynamically
moving nodes self-organize from a disconnected state to a surely
globally connected one. For our triangular lattice model,
$\sigma_{c}=0.37, 0.21, 0.13, 0.09$ and $0.065$ for
$r=2r_{0},3r_{0},4r_{0},5r_{0}$ and $6r_{0}$, respectively.
Obviously, various critical values of node occupancy $\sigma_{c}$
are required to ensure global connection for different transmission
ranges in MAHCN, which means that we can also inversely choose
proper transmission range to minimize energy consumption of the
network for different density of nodes on the lattice.

   It is natural to rescale $\eta(\sigma,r)$ into a universal
scaling function from direct observation of Fig. 2. we have
\begin{equation}
\eta \sim f(R^{\beta}\sigma)
\end{equation}
with reduced transmission range $R=(r-r_0)/r_0$.  the scaling index
$\beta$ of the universal function $f(x)$, respectively. We draw
transition curves in Fig.3 to show rescaling process. We can see
calculated curves for all $r$ collapse into the one with $r=2r_0$,
and the index $\beta=-0.49$ gives perfect convergence of all the
curves. This provides the evidence that the transitions at
$\sigma_c(r)$ are really critical phenomena. The inset of Fig.3
shows $\eta$ versus $\sigma$ for different sizes of MAHCN with
$r=2r_0$. The critical value $\sigma_c$ is independent of the sizes
of lattices, which is also valid for different transmission ranges.
Near critical points $\sigma_c(r)$, nodes with autonomic
communication self-organize into time-varying complex networks which
are reminiscent of directed dynamic small-world network (DDSWN)
model\cite{Zhu22} but with different scaling variables. Indeed, the
ratio of the number of nodes receiving message to the total number
of nodes should vary in the same way as that model provided the
nodes move at the same speed, have uniform transmission range and
relay message without delay, which will be discussed under another
title. But in the present work we check global connectivity with
burning algorithm, assuming that "combustion" (similar to message
spreading out) is much faster than variation of topological
structure. Moreover, the feature of transmission range-dependence
distinguishes itself from DDSWN model. It is also noticeable that Hu
and Chen\cite{HuJPA23} investigated scaling functions for bond
random percolation on honeycomb lattices with different aspect
ratios. By comparison, the present model is pertaining to maximally
connected network checked with combustion algorithm on a triangular
lattice although we always pay attention to the hexagonal cell
within the circular communication range.

\begin{center}
\begin{figure}
\scalebox{0.8}[0.7]{\includegraphics{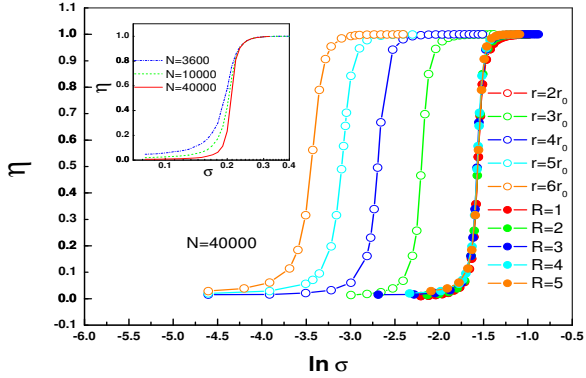}} \caption{(Colour
online)Scaling behavior of critical global connectivity of mobile
nodes for different transmission ranges. $\eta(\ln\sigma)$ curves
collapse into the same one with $r=2r_0$. Inset: Critical
connectivity under $r=2r_0$ with different sizes of the lattice.}
\end{figure}
\end{center}

    When the occupancy of nodes just exceeds the value of $\sigma_c$  for
certain transmission range $r$, they are found to self-organize into
a dynamically stationary network with our simulations. We can
characterize the connection of an ad hoc network with parameters of
complex networks. The simplest and the most intensively studied
parameter is degree distribution $p(k)$\cite{Barabasi} because it
governs fundamental properties of the system. Degree $k$ of a node,
as well known, is the total number of its topological edges
connecting with others. The dispersion of node degree is
characterized by the distribution function $p(k)$ which gives the
probability that a randomly selected node has exactly $k$ edges. In
the present paper, we study parameters of the network when it
consists of almost all the nodes of MAHCN together. The degree
distribution $p(k)$ follows Poisson distribution which is different
from that in ref.\cite{Xie6,Sarshar7,Sarshar8}. Fig.4 shows the
ensamble averaged degree distributions for all simulated
transmission ranges $r$ when the global connectivity of nodes is
above 0.9995. A cut-off degree $k_{c}=17$ appears in it, which means
that the probability for any node to have degree $k>k_c$ is very
low. The averaged degree of the whole network can be obtained from
direct observation of Fig.1, that is,
\begin{equation}
<k>=3r(r+1)\sigma
\end{equation}
where $r=zr_0$ as mentioned above. We recognize MAHCN from this kind
of Poisson-like distribution as random networks or small-world
networks. By the way, the global connectivity $\eta$ is also a
single-variable function of average degree $<k>$ for certain $r$,
which is qualitatively like the behavior of component size $S$ in
ref.\cite{Nemeth9} since $\sigma$ is proportional to $<k>$ in such
cases.

\begin{center}
\begin{figure}
\scalebox{0.8}[0.7]{\includegraphics{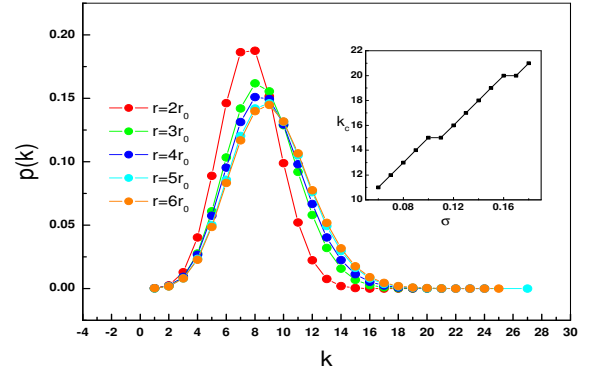}} \caption{(Colour
online) The degree distribution $p(k)$ of MAHCN for different
transmission ranges $r$. Inset: Cut-off degree $k_c$ versus varying
occupancy $\sigma$ for $r=4r_0$.}
\end{figure}
\end{center}

  Clustering coefficient $C$\cite{Watts} and $k_{nn}$\cite{Satorras},
the averaged nearest neighbor degree of nodes depicts complex
networks as viewed from correlations. For a node i, clustering
coefficient $C_i$ can be defined as the fraction of pairs of node
$i$'s neighbors that are also neighbors of each other in the
topological sense. $C(k)$ of the network is the clustering
coefficient averaged over nodes with the same value of degree $k$.
In our case, $C(k)=0.52, 0.55, 0.56, 0.57$, and $0.57$ for
transmission range $r=2r_0,3r_0,4r_0,5r_0$ and $6r_0$, respectively,
and keep invariant for $k\leq{k_c}$. The $k$-independent behavior of
$C(k)$ can be understood from the symmetry of hexagonal cells and
homogeneous distribution. Let us scrutinize the hexagonal cell
(orange in Fig.1) for $r=2r_0$ as an example. Site 1 in it has 18
neighbors (assuming full occupation). Therefore, the largest
possible number of closed topological triangles in the sense of
complex network should be $C_{18}^2$ which serves the denominator of
the clustering coefficient of it. Linking occupied sites 1 and 2, we
search for the third one with the distance less than $2r_0$ to both
of them in the hexagonal cell, which makes 7 triangles. Linking
occupied sites 3 and 4 to the center 1, makes 8 and 12 triangles,
respectively. Therefore, all equivalent sites
$((7+8+12)\times{6}/{2})$ in the cell make 81 closed triplets under
the constraint of maximum distance $2r_0$, and yields the clustering
coefficient of the site as ${81/C_{18}^2}=0.53$. Under the
assumption of homogeneous mixing, this is also valid for any
occupancy or averaged $<k>$ since we only need to multiply both the
numerator and the denominator of it by local $\sigma$
simultaneously, so that we have the same $C(k)$ for small
$k$($k\leq{k_c}$). The horizontal line for $k\leq{17}$ (see Fig. 5)
of $C(k)$ indicates particular constant clustering coefficient of
MAHCN for degrees occurring in high possibilities. Here value $k_c$
reflects the limit case of connection, and it may be pertinent to
the structure of the hexagonal cell imbedded in the triangular
lattice. The high value(above 0.5) of $C(k)$ implies that we have
dynamic small world networks at critical points, and there is large
redundancy in communication if only strategy of increasing occupancy
of sites is adopted. The averaged nearest neighbor degree of node
$i$ is simply: $k_{nn,i}={\sum\limits_{j}k_j}/{k_i}$, where $j$ is a
neighbor of node $i$. And $k_{nn}(k)$ can be accounted as the
function of degree k in the following form:
$k_{nn}(k)={\sum\limits_{k_{i}\in V}k_{nn,i}}/{\sum\limits_{k_{i}\in
V}1}$, where $V$ is the subset of nodes with the same degree $k$.
From figure 6 we can see that the curve of $k_{nn}$ versus $k$
suggests an empirical formula in the linear form for our MAHCN, that
is
\begin{equation}
k_{nn}(k)=b(r)+C(k)k
\end{equation}
where $b(r)$ is a k-independent constant. The positive assortativity
(i.e. increase behavior of $k_{nn}(k)$) gives the particular feature
distinguishing it from most other technical
networks\cite{Newman24,Li25,Vazquez26}. The appearance of tails in
large degrees $(k>17)$ (see Fig.5 and Fig.6) is due to very low
probabilistic occurrence in the simulation on 300 realizations.

\begin{center}
\begin{figure}
\scalebox{0.8}[0.7]{\includegraphics{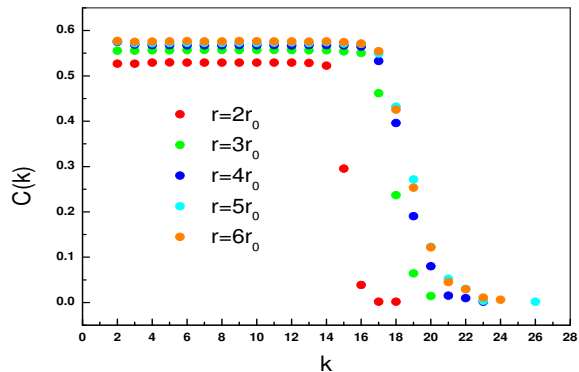}} \caption{(Colour
online)The clustering coefficient $C(k)$ for different transmission
ranges $r$ on the triangular lattice.}
\end{figure}
\end{center}

\begin{figure}
\scalebox{0.8}[0.7]{\includegraphics{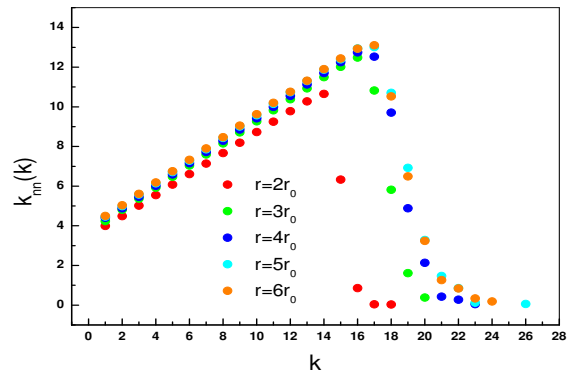}} \caption{(Colour
online)The averaged nearest neighbor degree of nodes $k_{nn}(k)$
versus degree $k$ for different transmission ranges $r$ on the
triangular lattice.}
\end{figure}

   Interestingly, the model can be thought as a special example for the
process of coevolutionary competitive exclusion\cite{Zhu27}: taking
sites as the state variable of moving nodes, each node tends to
compete for links to others, which is demanded by the global
connection. Only nodes within other ones' transmission ranges and
with the degrees of both ones less than $k_c$ can successfully gain
links, which constitutes threshold conditions of the co-evolution
network.

  In summary, we present a model for mobile ad hoc communication
networks by considering uniform transmission range of nodes and
assigning moving nodes randomly on the plane of the triangular
lattice. Demanded by the best communication, critical global
connectivity is found from simulations for various transmission
ranges by adjusting the node occupancy of sites on the lattice. The
order parameter scales with transmission ranges, but behaves
differently from other models. This forms a kind of percolation in
dynamic complex networks pertaining to geographic distance.
Moreover, cut-off degree $k_{c}$, invariant clustering coefficient
and linear assortativity as functions of degree are found for
self-organized complex networks near critical global connectivity.
Our model suggests that transmission range of nodes and average
occupancy on the plane should adapt to each other to balance
minimization of energy consumption with the global connectivity. In
fact, nodes relaying message consume energy continuously.
Transmission range of each node usually reduces with time relating
to its job load. Therefore, the present model is applicable only for
the routing strategy of broadcasting\cite{Jetcheva28}. A more
practical model should include random distributed nodes with
changing transmission ranges. Meanwhile, nodes are not necessary to
move along edges of the triangular lattice. Therefore, a random
graph model without any lattice is necessary for better
investigation on mobile ad hoc communication networks, which leaves
our further work in the future.

  We acknowledge partial support from the National Natural Science Foundation
of China (NNSFC) under the Grant Nos. 70471084, 10775071, 10635040
and 60676056. CPZ  and BHW thank the hospitable accommodations of
Bao-Wen Li in NUS. CPZ thanks Xiang-Tao Fan, Y.-C. Lai and B. J. Kim
for checking up English of the manuscript. We also acknowledge the
support by National Basic Science Program of China Project
Nos.2005CB623605, 2006CB921803 and 2006CB705500.

{}

\end{document}